\documentstyle[twocolumn,aps,prl,overcite]{revtex}

\input epsf

\begin{document}
\title{Essentials of $k$-Essence}
\author{C. Armendariz-Picon,$^1$ V. Mukhanov$^1$ and Paul J. Steinhardt$^{2}$}
\address{$^1$Ludwig Maximilians Universit\"{a}t, Sektion Physik, \\ 80333
M\"{u}nchen, Germany \\
$^2$Department of Physics, Princeton University, Princeton, NJ 08540}
\date{\today}
\maketitle

\begin{abstract}
We recently introduced the concept of ``$k$-essence" as a dynamical solution
for explaining naturally why the universe has entered an epoch of
accelerated expansion at a late stage of its evolution. The solution avoids
fine-tuning of parameters and anthropic arguments. Instead, $k$-essence is
based on the idea of a dynamical attractor solution which causes it to act
as a cosmological constant only at the onset of matter-domination.
Consequently, $k$-essence overtakes the matter density and induces cosmic
acceleration at about the present epoch. In this paper, we present the basic
theory of $k$-essence and dynamical attractors based on evolving scalar
fields with non-linear kinetic energy terms in the action. We present
guidelines for constructing concrete examples and show that there are two
classes of solutions, one in which cosmic acceleration continues forever and
one in which the acceleration has finite duration.
\end{abstract}

\section{Introduction}

A concordance of cosmological observations\cite{Bahcalletal} of large-scale
structure, the cosmic microwave background anisotropy and Type IA supernovae
at deep red shift suggest that the matter density of the universe comprises
about one-third of the critical value expected for a flat universe. The
missing two-thirds is due to an exotic dark energy component with negative
pressure that causes the Hubble expansion to accelerate today. One candidate
for such a component is a cosmological constant ($\Lambda $) or vacuum
density. Another possibility is a dynamical component whose energy density
and spatial distribution evolve with time, as is the case for quintessence 
\cite{CaDaSt} or, as explored herein, for $k$-essence.\cite{ArMuSt}

A key challenge for theoretical physics is to address the cosmic coincidence
problem: why does the dark energy component have a tiny energy density ($%
{\cal O}({\rm meV}^4)$) compared to the naive expectation based on quantum
field theory and why does cosmic acceleration begin at such a late stage in
the evolution of the universe. Most dark energy candidates (such as the
cosmological constant) require extraordinary fine-tuning of the initial
energy density to a value 100 orders of magnitude or more smaller than the
initial matter-energy density. Proponents of anthropic models\cite{Weinberg}
often pose the problem as: why should the acceleration begin shortly after
structure forms in the universe and sentient beings evolve? If the dark
energy component consists of vacuum density ($\Lambda$) or quintessence\cite
{CaDaSt} in the forms that have been discussed in the literature to date,
the answer is either pure coincidence or the anthropic principle.

The purpose of introducing $k$-essence is to provide a dynamical explanation
which does not require the fine-tuning of initial conditions or mass
parameters and which is decidedly non-anthropic. In this scenario, cosmic
acceleration and human evolution are related because both phenomena are
linked to the onset of matter-domination. The $k$-essence component has the
property that it only behaves as a negative pressure component after
matter-radiation equality, so that it can only overtake the matter density
and induce cosmic acceleration after the matter has dominated the universe
for some period, at about the present epoch. And, of course, human evolution
is linked to matter-domination because the formation of planets, stars,
galaxies and large-scale structure only occurs during this period. A further
property of $k$-essence is that, because of the dynamical attractor
behavior, cosmic evolution is insensitive to initial conditions.

The existence of attractor solutions is reminiscent of quintessence models
based on evolving scalar fields with exponential\cite{expon} ``tracker"\cite
{ZlWaSt,StWaZl} potentials. In these models, an attractor solution causes
the energy density in the scalar field to track the equation-of-state of the
dominant energy component, be it radiation or matter. An advantage is that
the cosmic evolution is insensitive to the initial energy density of the
quintessence field, and, for many models,
the scenario can begin with the most natural
possibility, equipartition initial conditions. (For the case of vacuum
energy or cosmological constant, the vacuum energy must be set 120 orders of
magnitude less than the initial matter-radiation density.) However, so long
as the field tracks any equation-of-state, it cannot overtake the
matter-density and induce cosmic acceleration. Indeed, for a purely
exponential potential, the field never overtakes the matter density and 
dominates the universe. Hence,  this
is an unacceptable candidate for the dark energy component. In tracker
models, the problem is addressed because the curvature of the potential
ultimately dips to a critically small value once the field passes a
particular value, $\bar{Q}$ such that the field $Q$ becomes frozen and
begins to act like a cosmological constant. The value of the potential
energy density at $Q=\bar{Q}$ determines when quintessence overtakes the
matter density and cosmic acceleration begins. The overall scale of the
potential must be finely adjusted in order for the component to overtake the
matter-density at the present epoch. So, while tracker models allow
equipartition initial conditions, they require the same fine-tuning as
models with cosmological constant.

The distinctive feature of the $k$-essence models we consider is that
$k$-essence only tracks the equation-of-state of the background during the
radiation-dominated epoch. A tracking solution during the matter-dominated
epoch is physically forbidden. Instead, at the onset of
matter-domination, the $k$-essence field energy density $\varepsilon $ drops
several orders of magnitude as the field approaches a new attractor solution
in which it acts as a cosmological constant with pressure $p$ approximately
equal to $-\varepsilon $. That is, the equation-of-state, $w\equiv
p/\varepsilon $, is nearly -1. The $k$-essence energy density catches up and
overtakes the matter-density, typically several billions of years after
matter-domination, driving the universe into a period of cosmic
acceleration. As it overtakes the energy density of the universe, it begins
to approach yet another attractor solution which, depending on details, may
correspond to an accelerating universe with $w<-1/3$ or a decelerating or
even dust-like solution with $-1/3<w\leq 0$. In this scenario, we observe
cosmic acceleration today because the time for human evolution and the time
for $k$-essence to overtake the matter density are both severals of billions
of years due to independent but predictive dynamical reasons.

The $k$-essence models which we have found rely on dynamical attractor
properties of scalar fields with non-linear kinetic energy terms in the 
action, models which are unfamiliar to
most particle physicists and cosmologists. Some of the concepts were first
introduced to develop an alternative inflationary model known as
$k$-inflation.\cite{ArDaMu} In this paper, we present a thorough, pedagogical
study of dynamical attractor behavior and the application to present-day
cosmic acceleration. The paper is organized as follows: In Section II we
derive the basic equations describing the dynamics of a universe filled by
matter, radiation and $k$-essence. In Section III, we classify the possible
attractor solutions for $k$-essence. In some cases, the attractor solution
causes $k$-essence to mimic the equation-of-state of the background energy
density; we refer to this as a {\it tracker} solution. In other cases,
$k$-essence mimics a cosmological constant, quintessence or dust 
without depending on the presence of
any additional cosmic energy density. In Section IV, we show how these
principles can be used to control how $k$-essence travels through a series
of attractor solutions as the universe evolves beginning from general
initial conditions. In particular, we show how $k$-essence can transform
automatically into an effective cosmological constant at the onset of
matter-domination, as is desired to explain naturally the present-day cosmic
acceleration. In Section V, we show how to utilize these concepts
to design model Lagrangians.  We explore 
two 
illustrative examples. In one case, the future evolution of $k$-essence
causes the universe to accelerate forever. In the other case, $k$-essence
ultimately approaches an equation-of-state corresponding to pressureless
dust, and the universe returns to a decelerating phase.

\section{Basics of $k$-essence}

The attractor behavior required for avoiding the cosmic coincidence problem
can be obtained in models with non-standard (non-linear) kinetic energy
terms. In string and supergravity theories, non-standard kinetic terms
appear generically in the effective action describing the massless scalar
degrees of freedom. Normally, the non-linear terms are ignored because they
are presumed to be small and irrelevant. This is a reasonable expectation
since the Hubble expansion damps the kinetic energy density over time. However,
one case in which the non-linear terms cannot be ignored is if there is an
attractor solution which forces the non-linear terms to remain
non-negligible. This is precisely what is being considered here. Hence, we
wish to emphasize that $k$-essence models are constructed from building
blocks that are common to most quantum field theories and, then, utilize
dynamical attractor behavior (that often arises in models with non-linear
kinetic energy) to produce novel cosmological models.

Restricting our attention to a single field, the action generically may be
expressed (perhaps after conformal transformation and field redefinition) as 
\begin{equation}
S_{\varphi }=\int d^{4}x\,\sqrt{-g}\left[ -\frac{R}{6}+p(\varphi ,X)\right] ,
\label{eq:action}
\end{equation}
where we use units such that $\frac{8\pi G}{3}=1,$ and 
\begin{equation}
X=\frac{1}{2}(\nabla \varphi )^{2}.  \label{eq:X}
\end{equation}
The Lagrangian $p$ depends on the specific particle theory model. In this
paper, we consider only factorizable Lagrangians of the form 
\begin{equation}
p=K(\varphi )\widetilde{p}(X),  \label{eq:p}
\end{equation}
where we assume that $K(\varphi )>0.$

Lagrangians of this type are general enough to accommodate slow-roll,
power-law and pole-like inflation, and they also appear rather naturally in
the effective action of string theory. For small $X$, one can have $%
\widetilde{p}(X)=const.+X+{\cal O}(X^{2})$. Ignoring quadratic and higher
order terms, the theory corresponds (after field redefinition) to an
ordinary scalar field with some potential. Normally, higher order kinetic
energy terms are ignored under the assumption that they are small, but the
attractor solutions considered here insure that the non-linear terms remain
non-negligible throughout cosmic history. The scalar field for which these
higher order kinetic terms play an essential role we call, for brevity,
$k-$essence.

To describe the behavior of the scalar field it is convenient to use a
perfect fluid analogy. The role of the pressure is played by the Lagrangian
$p$ itself, while the energy density is given by \cite{ArDaMu} 
\begin{eqnarray}
\varepsilon  &=&K(\varphi )(2X\widetilde{p}_{,X}(X)-\widetilde{p}(X))
\label{eq:energy} \\
&\equiv &K(\varphi )\tilde{\epsilon}(X),
\end{eqnarray}
where $..._{,X}$ denotes a partial derivative with respect to $X$. The ratio
of pressure to energy density, which we call, for brevity, the $k$-essence
equation-of-state, 
\begin{equation}
w_{k}\equiv \frac{p}{\varepsilon }=\frac{\widetilde{p}}{\tilde{\epsilon}}=
\frac{\widetilde{p}}{2X\widetilde{p}_{,X}-\widetilde{p}},  \label{eq:w-k}
\end{equation}
does not depend on the function $K(\varphi )$. For a ``standard'' kinetic
term, $p=X$, in the case when there is no potential, the equation-of-state
is $w_{k}=1$. However, for a general choice of $p$ it is easy to get any
value of $w_{k}$. Notice that $w_{k}<-1$ does not imply necessarily the
instability of the fluid with respect to small wavelength perturbations. The 
{\it effective} ``speed of sound,'' $c_{S}$, which determines the
propagation of perturbations in the $k$-essence component is \cite{GaMu} 
\begin{equation}
c_{S}^{2}=\frac{p_{,X}}{\varepsilon _{,X}}=\frac{\widetilde{p}_{,X}}
{\tilde{\epsilon}_{,X}},  \label{eq:cssq}
\end{equation}
and it can be positive for any $w_{k}$. For instance, the effective speed of
sound, defined to be the coefficient of the momentum-squared term in the
perturbation equation for the scalar field, is always equal one for
quintessence models with canonical kinetic energy, while the equation of
state $w$ can be rather arbitrary here.

We want to study the evolution of a universe filled by $k$-essence (labeled
in the eqs. below by ``k'') and matter-radiation (labeled by ``m'' in cases
where we refer generically to the dominant matter-radiation component, be it
dust-like or radiation, or by ``d'' or ``r'' if we refer specifically to the
dust-like or radiation component, respectively). There is increasing
evidence that the total energy density of the universe is 
equal to the critical value,\cite{Bahcalletal,boom} and,
hence,  we will consider a flat universe only. In that case the equation for
the scale factor $a$ takes the form 
\begin{equation}
H^{2}\equiv \dot{N}^{2}=\varepsilon _{m}+\varepsilon _{k},
\label{eq:hubble-constant}
\end{equation}
where a dot denotes derivative with respect to physical time $t$ and we
introduced the number of e-foldings $N=\log a$. This equation has to be
supplemented with equations for $\varepsilon _{m}$ and $\varepsilon_{k}$.
These are the energy conservation equations for each component $j$: 
\begin{equation}
\frac{d\varepsilon _{j}}{dN}=-3\varepsilon _{j}(1+w_{j}),
\label{eq:$k$-motion}
\end{equation}
where $w_{j}$ is the equation-of-state for the appropriate matter-radiation
or $k$-essence component. Considering a homogeneous field $\varphi $ and
substituting the expression for the energy density (\ref{eq:energy}) into
the appropriate Eq.~(\ref{eq:$k$-motion}) for $k$-essence, one gets 
\begin{equation}
\frac{dX}{dN}=-\frac{\tilde{\epsilon}}{\tilde{\epsilon}_{,X}}\left[
3(1+w_{k})+\sigma \frac{K_{,\varphi }}{K}\frac{\sqrt{2X}}{H}\right],
\label{eq:master}
\end{equation}
where $w_{k}$ is given by Eq.~(\ref{eq:w-k}), $\sigma \equiv {\rm sign}
(d\varphi /dN)$ and the Hubble constant is given by Eq.~(\ref
{eq:hubble-constant}).

We will consider functions $\widetilde{p}\left( X\right)$ 
that increase monotonically with  $X$. They
should satisfy further restrictions, which follow from the requirements of
positivity of the energy density, 
\begin{equation}
\tilde{\epsilon}=2X\widetilde{p}_{,X}-\widetilde{p}>0  \label{posenergy}
\end{equation}
and stability of the $k$-essence background, $c_{S}^{2}>0,$ implying 
\begin{equation}
\tilde{\epsilon}_{,X}=2X\widetilde{p}_{,XX}+\widetilde{p}_{,X}>0.
\label{posspeed}
\end{equation}

For designing models and visualizing constraints, it is helpful to
re-express $\widetilde{p}$ as $\widetilde{p}=g\left( y\right) /y$ and
consider it as a function of the  new variable $y=X^{-1/2}.$ The pressure of
the $k$-essence component is, therefore, 
\begin{equation}
p=K(\varphi )g(y)/y;  \label{press}
\end{equation}
the equation-of-state and the effective sound speed are, correspondingly, 
\begin{equation}
w_{k}=\frac{p}{\varepsilon }=-\frac{g}{yg^{\prime }};\,\,{\rm \ \ \ \ \ \ \
\ \ \ \ \ }\,\,c_{S}^{2}=\frac{(g-g^{\prime }y)}{g^{\prime \prime }y^{2}}
\end{equation}
and the restrictions Eq.~(\ref{posenergy}) and (\ref{posspeed}) take the
very simple form 
\begin{equation}
\tilde{\epsilon}=-g^{\prime }>0,\text{ \ \ \ }\tilde{\epsilon}_{,X}=\frac{1}
{2}y^{3}g^{\prime \prime }>0,\text{\ \ \ \ \ \ \ \ \ }  \label{newenergy}
\end{equation}
where prime denotes derivative with respect to $y$. These conditions just
mean that $g$ should be a decreasing convex function of $y=X^{-1/2}$. A
generic function which satisfies these restriction is shown in Fig. 1.
Taking into account that $H=\sqrt{\varepsilon _{tot}}=\sqrt{\varepsilon
_{m}+\varepsilon _{k}}$ and $\varepsilon _{k}=K\left( \varphi \right)
\tilde{\epsilon}\left( y\right) =-Kg^{\prime }\left( y\right)$ one can rewrite
Eq.~(\ref{eq:master}) in terms of the new variables as 
\begin{equation}
\frac{dy}{dN}=\frac{3}{2}\frac{(w_{k}\left( y\right) -1)}{r^{\prime }\left(
y\right) }\left[ r\left( y\right) +\sigma \frac{K_{,\varphi }}{2K^{3/2}}
\sqrt{\frac{\varepsilon _{k}}{\varepsilon _{tot}}}\right] ,  \label{master0}
\end{equation}
where 
\begin{equation}
r\left( y\right) \equiv (-\frac{9}{8}g^{\prime })^{1/2}\,y\,\left( 1+w_{k}
\right) =\frac{3}{2\sqrt{2}}\frac{\left( g-g^{\prime }y\right)}
{\sqrt{-g^{\prime}}}
\label{definitionr}
\end{equation}
is a  function which, as we will see later, is  critical for  the attractor
properties of $k-$essence.

\begin{figure}[tbp]
\begin{center}
 \epsfxsize=3.3 in \epsfbox{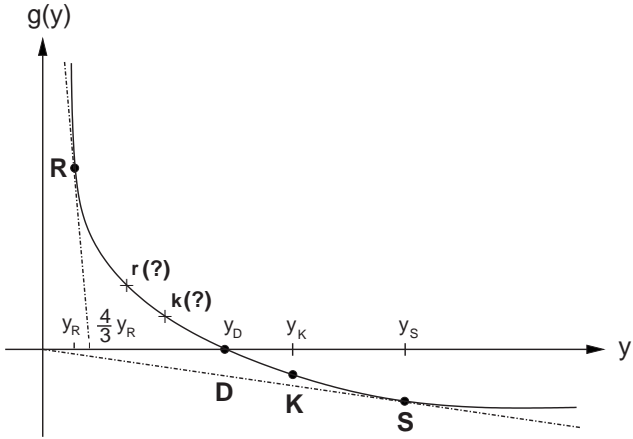}
 \end{center}
\caption{A sample function $g(y)$. Boldface letters denote the corresponding
attractors; their positions are given on the $y$-axis. The tangent to the
curve at a radiation tracker, such as ${\bf R}$,
goes through $4y_R/3$, whereas the
tangent to the curve at the de Sitter point ${\bf S}$ goes through the
origin.}
\end{figure}

\section{Classification of Tracker and Attractor Solutions}

The attractor solutions for $k$-essence can be divided into two classes. In
one class, $k$-essence mimics the equation-of-state of the matter-radiation
component in the universe. We refer to these  as {\it %
trackers} because the cosmic evolution of $k$-essence
follows the track of another energy component.  
The second class of attractors consists of cases where $k$-essence
is drawn towards an equation-of-state which is different from  matter or
radiation. These attractors are important in the limits
where  $k$-essence is either a negligibly  small or an overwhelming large 
fraction of the total energy density.
The types of attractors available at any given moment in cosmic history depend
on whether the universe is radiation- or matter-dominated. 
For all types of attractors,
there is an associated basin of attraction, a set of initial conditions
which evolve towards the attractor.

In the presence of a matter background (dust or radiation) component with
constant equation-of-state $w_{m}$, Eq.~(\ref{master0}) can have tracking
solutions for which the $k$-essence equation-of-state 
equals  $w_{m}.$ To reveal when it can happen and to find these
solutions explicitly we just need to note that if such solutions exist,
they have to be generically of the form $y\left( N\right) =y_{m}=const.,$
where $y_{m}$ satisfies the equation 
\begin{equation}
w_{k}\left( y_{m}\right) \equiv \left. -\frac{g}{yg^{\prime }}\right|
_{y=y_{m}}=w_{m}.  \label{attractoreq}
\end{equation}
Substituting this ansatz into Eq.~(\ref{master0}) and noting that the ratio
$\varepsilon _{k}/\varepsilon _{tot}$ should stay constant during the
tracking stage, we see that $y\left( N\right) =y_{m}$ can be a solution of
Eq.~(\ref{master0}), only if $K\left( \varphi \right) =const./\varphi ^{2}$
and, therefore, for simplicity,
we consider from now on only scalar fields with Lagrangian 
\begin{equation}
L=\frac{g\left( y\right) }{\varphi ^{2}y}.  \label{finallagrangian}
\end{equation}
It is worth noting that this kind of  dependence on a scalar field occurs in
the string tree-level effective action when expressed in the Einstein
frame.\cite{Witten,Lukas1,Lukas2}  In this case, Eq.~(\ref{master0})
simplifies to 
\begin{equation}
\frac{dy}{dN}=\frac{3}{2}\frac{(w_{k}\left( y\right) -1)}{r^{\prime }\left(
y\right) }\left[ r\left( y\right) -\sqrt{\frac{\varepsilon _{k}}{\varepsilon
_{tot}}}\right] ,  \label{master1}
\end{equation}
where we restrict ourselves to the most interesting case of positive $\sigma 
$ on the branch of positive $\varphi .$ To close the system of equations for
the two unknown variables $y$ and $\varepsilon _{k}/\varepsilon _{tot},$ we
use the equation 
\begin{equation}
\frac{d\left( \varepsilon _{k}/\varepsilon _{tot}\right) }{dN}=3\frac
{\varepsilon _{k}}{\varepsilon _{tot}}\left( 1-\frac{\varepsilon _{k}}
{\varepsilon _{tot}}\right) \left( w_{m}-w_{k}\left( y\right) \right) ,
\label{master1a}
\end{equation}
which immediately follows from Eq.~(\ref{eq:$k$-motion}). If $y_{m}$ is a
solution of Eq.~(\ref{attractoreq}), 
then $%
y\left( N\right) =y_{m}=const,$ satisfies Eqs.~(\ref
{master1}) and (\ref{master1a}), provided 
\begin{equation}
r^{2}\left(y_{m}\right)=
\left( \frac{\varepsilon _{k}}{\varepsilon _{tot}}\right) _{m}<1,  
\label{ratio}
\end{equation}
where the inequality is simply the physical constraint that 
$\varepsilon_{k}< {\varepsilon _{tot}}$ (assuming  positive energy densities
$\varepsilon_{k}$ and $\varepsilon _{m}$).
If $r\left( y_{m}\right) >1,$ a tracker solution $y(N)=y_{m}$
is physically forbidden.

\subsection{When are trackers attractors?}

To find out when trackers are stable solutions with a non-trivial basin of
attraction, we  study the behavior of small deviations from the tracker
solution.
Substituting $y\left( N\right) =y_{m}+\delta y$ and $\varepsilon
_{k}/\varepsilon _{tot}\left( N\right) =(\varepsilon _{k}/\varepsilon
_{tot})_{m}+\delta (\varepsilon _{k}/\varepsilon _{tot})$ into Eqs.~(\ref
{master1}) and (\ref{master1a}) and linearizing, 
we obtain 
\begin{equation}
\frac{d\delta y}{dN}=\frac{3}{2}\frac{(w_{k}\left( y_{m}\right) -1)}
{r_{m}^{\prime }}\left[ r_{m}^{\prime }\delta y-\frac{\delta (\varepsilon
_{k}/\varepsilon _{tot})}{2r_{m}}\right] ,  \label{lin1}
\end{equation}
\begin{equation}
\frac{d\delta \left( \varepsilon _{k}/\varepsilon _{tot}\right) }{dN}
=-3r_{m}^{2}\left( 1-r_{m}^{2}\right) w_{k}^{\prime }\left( y_{m}\right)
\delta y,  \label{lin2}
\end{equation}
where the index $``m"$ denotes evaluation of the appropriate quantities at
the tracker point $y_{m}$ and $(\varepsilon _{k}/\varepsilon _{tot})_{m}$
has been replaced by $r^{2}\left( y_{m}\right) $ according to Eq.~(\ref
{ratio}). Differentiating Eq.~(\ref{lin1}) with respect to $N$ and using
Eq.~(\ref{lin2}), one obtains the following closed equation for $\delta y$: 
\begin{equation}
\frac{d^{2}\delta y}{dN^{2}}+\frac{3}{2}\left( 1-w_{m}\right) \frac{d\delta y}
{dN}+\frac{9}{2}(1-r_{m}^{2})(1+w_{m})(c_{S}^{2}-w_{m})\delta y=0.
\label{lin3}
\end{equation}
Here $c_{S}^{2}$ is the squared ``speed of sound'' of $k$- essence at the
tracker point and we took into account that $w_{k}\left( y_{m}\right) =w_{m}$.
Eq.~(\ref{lin3}) is a second order differential equation with constant
coefficients and has two exponential solutions. It is easy to see that for
$\left| w_{m}\right| <1$ both solutions decay if 
\begin{equation}
c_{S}^{2}>w_{m}.  \label{stabcond}
\end{equation}
Therefore, since $c_{S}^{2}=(g-g^{\prime }y)/g^{\prime \prime }y^{2}$, any
tracker can be easily made an attractor by arranging a small second
derivative of $g$ at the tracker point.

As important examples, let us consider the two most interesting cases,
namely, trackers in the presence of radiation (labeled  ``r'' in the
equations below) and cold matter (labeled ``D'' for  ``dust'').

\subsection{ Radiation trackers}

For radiation trackers, $w_{m}\equiv w_{r}=1/3$ and Eq.~(\ref{attractoreq}),
which defines the location of the radiation trackers ($y_{m}\equiv y_{r}$),
reduces to 
\begin{equation}
y_{r}g^{\prime }(y_{r})=-3g(y_{r}).  \label{trarad}
\end{equation}
The ratio of the energy densities is given by 
\begin{equation}
\left( \frac{\varepsilon _{k}}{\varepsilon _{tot}}\right) _{r}=r^{2}\left(
y_{r}\right) \equiv -2g^{\prime }(y_{r})y_{r}^{2}  \label{ratiorad}
\end{equation}
and radiation trackers exist only if at the points $y_{r}$ satisfying Eq.
(\ref{trarad}), $r^{2}\left( y_{r}\right) <1$. These trackers are stable
attractors only if $g^{\prime \prime }\left( y_{r}\right) <-4g^{\prime
}\left( y_{r}\right) /y_{r}.$ Radiation trackers are always located in the
region where $g>0$ (positive pressure), corresponding to $y<y_{D}$ in Fig.
1.  For a given $g(y)$, there can be more than one radiation 
tracker. For each of them, the geometrical way of finding the 
value of $y$ corresponding to the tracker is given in Fig.~1.
These trackers can have different values
of $r^2(y_r) = \left( \varepsilon _{k}/\varepsilon _{tot}\right) _{r}$. 
 Numerically,
a likely range for $r^2(y)$ is
$10^{-1} - 10^{-2}$. This is also the range we wish to have in 
order that cosmic acceleration begin at roughly the present epoch.
We label the radiation tracker  with the desired value of $r^2(y_R)$ as
${\bf R}$, and a  second possible radiation
tracker with a different value of $r^2 (y_r)$ (the one closest to $y_D$) 
as ${\bf r}$(?) in Fig.~1. If $r^2(y_r)$ is much smaller than $10^{-2}$, 
the energy density falls so much at the onset
of matter-domination (before it freezes at a constant value) that it 
would not yet have overtaken the matter density today.  
If $r^2(y_r)$ is much greater than $10^{-1}$, then the 
contribution of $k$-essence to the total energy density would
change the expansion rate in the early universe 
and adversely affect the predictions of primordial nucleosynthesis. 
The current constraints on $r^2(y_r)$ from nucleosynthesis vary
from 4 per cent\cite{Tytler} to 20 per cent\cite{Olive}, 
depending on how the observations are weighted.

\subsection{Dust trackers}

The $k$-essence field can also track the dust ($w_{D}=0)$ in the (cold)
matter-dominated universe$.$ Since the pressure 
is proportional to $g(y)$ and  is
zero for  dust, it must be that 
\begin{equation}
g\left( y_{D}\right) =0,  \label{attrdust}
\end{equation}
at the dust attractor point, $y=y_{D}$. An additional condition for
the existence of the dust tracker is that $r\left( y_{D}\right) <1$ (see
discussion following Eq.~(\ref{ratio})). In this case the ratio of energy
densities at the dust tracker is given by 
\begin{equation}
\left( \frac{\varepsilon _{k}}{\varepsilon _{tot}}\right) _{D}=r^{2}\left(
y_{D}\right) =-\frac{9}{8}\,g^{\prime }(y_{D})\,y_{D}^{2}.  \label{ratiodust}
\end{equation}
If a dust tracker exists then it is always an attractor, since the stability
condition Eq.~(\ref{stabcond}) just means here that the ``speed of sound''
of $k-$essence should be positive. Note, that for the 
monotonically decreasing convex
functions $g$ under consideration only a maximum of one dust attractor can
exist (see Fig.1) since $g$ has only one zero. It is very important to point
out that one can easily avoid a dust tracker by considering functions $g$
such that $r^{2}\left( y_{D}\right) =-\frac{9}{8}\,g^{\prime
}(y_{D})\,y_{D}^{2}>1$ at $y_{D}$.

\subsection{De Sitter Attractors}

We have noted  that $k$-essence  can have 
 attractor solutions which are not trackers 
 in that they do not mimic matter or radiation. These
attractor solutions play an important role in two extreme cases, namely,
when the energy density of matter or radiation is either much bigger or much
smaller than the energy density of $k$-essence. In this subsection, we study
the case when the background is dominated by matter-radiation and
$k$-essence is an insignificant component, $\varepsilon _{k}\ll \varepsilon
_{m}.$ In this case, if $g(y)$ satisfies some simple properties, $k$-essence
has an attractor solution in which it behaves like a cosmological constant
($w_{k}\rightarrow -1$). We refer to this solution as the de Sitter attractor
(labeled ``{\bf S}'').

Our purpose is to construct models in which $k$-essence has a 
positive pressure, radiation
tracker solution ({\bf R}) during the radiation-dominated phase 
and approaches a state with negative pressure
shortly  
after the onset
of the matter-dominated phase. 
At the very least, it is necessary that
$g(y)$ be positive for some range of $y$ and negative for another range
since the pressure is proportional to $g(y)$.  
This 
simple
condition is generically
sufficient to produce  a de Sitter attractor solution:
Since $g^{\prime }$ must be negative (the positive energy condition, Eq.
(\ref{newenergy})), it follows that $g$ must have a unique zero, $y_{D}$, the
only dust attractor possible. Furthermore, $g(y)$ is positive for $y<y_{D}$,
a  range which must include the radiation tracker, $y=y_{R}$. For $y>y_{D}$,
the pressure ($\propto g$) and, correspondingly, $w_{k}=-g/yg^{\prime }$ are
negative. From this observation, combined with 
the stability condition ($g^{\prime
\prime }>0$; see Eq.~(\ref{newenergy})), 
it follows that the derivative of $%
r(y)$ (see definition~(\ref{definitionr})) 
\begin{equation}
r^{\prime }=\frac{3}{4\sqrt{2}}\frac{g^{\prime \prime }y}{\sqrt{-g^{\prime }}}
\left( w_{k}-1\right)   \label{rderivative}
\end{equation}
must be negative for $y>y_{D}$. Since $r\left( y\right) $ is positive at
$y=y_{D}$ and has a negative derivative for $y>y_{D},$ generically (provided
$r^{\prime }$ does not approach zero too rapidly) $r(y)$ should vanish at
some point $y=y_{S}$ $>y_{D}$ and then become negative. As immediately
follows from the definition of $r$ (see (\ref{definitionr})), the
equation-of-state of $k$-essence at $y=y_{S}$ (point $S$ in Fig.1)
corresponds to a cosmological term: $w_{k}(y_{S})=-1$.
 Hence, we see that de Sitter attractors exist for a 
very wide class of $g(y)$ and are a generic feature of $k$-essence 
models.

In the absence of matter, $y\left( N\right) =y_{S}=const.$ is not a solution
of the equations of motion. However, when matter strongly dominates over
$k$-essence ($\varepsilon _{k}/\varepsilon _{tot}\ll 1)$, there exists a
solution in the vicinity of this point. (Formally,
in the limit $\varepsilon _{k}/\varepsilon
_{tot}\rightarrow 0,$ $y\left( N\right) \rightarrow 
y_{S}$ is an exact solution of
Eqs.~(\ref{master1}) and (\ref{master1a}).) 
Setting $w_{m}=w_{k}=-1$ in Eq.~(\ref{lin1}) it can be also verified that
this is a stable attractor. For finite,
but very small ratio $\varepsilon _{k}/\varepsilon _{tot}\ll 1,$ the
approximate solution, corresponding to $w\approx -1,$ is located in the
vicinity of $y_{S}$ and has the form: 
\begin{equation}
\frac{\varepsilon _{k}}{\varepsilon _{tot}}\left( N\right) \propto \exp
\left( 3\left( 1+w_{m}\right) N\right)   \label{desitter1}
\end{equation}
and
\begin{equation}
y\left( N\right) \approx y_{S}+\frac{2}{\left( 3+w_{m}\right) r^{\prime
}\left( y_{S}\right) }\left( \frac{\varepsilon _{k}}{\varepsilon _{tot}}
\left( N\right) \right) ^{1/2}. \label{desitter2}
\end{equation}

As shown below, if at any moment of time $%
\varepsilon _{k}/\varepsilon _{tot}$ lies below the basin of attraction of
the tracker solutions, $k$-essence will be driven first to the de Sitter
attractor and stay in its vicinity as long as $\varepsilon _{k}/\varepsilon
_{tot}$ is sufficiently small. We will utilize this property at the
transition from the radiation- to the matter-dominated phase.

\subsection{$k$-Attractors}

Whereas the de Sitter attractors are important when $k$-essence is an
insignificant contribution to the total energy density, the $k$-attractors
arise when $k$-essence is the dominant energy component. In the absence of
matter ($\varepsilon _{k}/\varepsilon _{tot}=1),$ the function $y\left(
N\right) =y_{k}=const,$ where $y_{k}$ satisfies the equation 
\begin{equation}
r\left( y_{k}\right) =1,  \label{kattractor}
\end{equation}
is a solution of Eq.~(\ref{master1}), while Eq.~(\ref{master1a}) is
satisfied identically. This solution describes a power-law expanding
universe.\cite{ArDaMu,GaMu} The equation-of-state can be easily
obtained from Eqs.~(\ref
{definitionr}) and (\ref{kattractor}): 
\begin{equation}
1+w_{k}\left( y_{k}\right) =\frac{2\sqrt{2}}{3}\frac{1}{\sqrt{-g_{k}^{\prime}
y_{k}^{2}}}=const,  \label{keqofstate}
\end{equation}
and the scale factor is 
\begin{equation}
a\propto t^{\frac{2}{3\left( 1+w_{k}\right) }}=t^{\sqrt{-g_{k}^{\prime
}y_{k}^{2}/2}}.  \label{scalefactor}
\end{equation}
If $-g_{k}^{\prime }\,y_{k}^{2}/2>1$ the solution describes power law
inflation, which is an attractor of the system provided that $r^{\prime}
\left( y_{k}\right) <0$.

The existence of a $k$-attractor depends mainly on the form of the function
$r(y)$. A  $k$-attractor corresponds to
$r(y_{k}) \rightarrow 1$ ({\it i.e.}, the limit where  the energy
density is totally dominated by $k$-essence). In general, if $r(y_{0})>1$ for
some $y_{0}$ and there exists an {\bf S}-attractor ($r(y_{S})=0$), then there
must exist a $k$-attractor somewhere between them, $y_{0}<y_{k}<y_{S}$,
simply because $r\left( y\right) $ is a continuous function. 

In particular,
we are interested in the case where there is no dust attractor because
$r(y_{D})>1$, and yet there is a de Sitter attractor with $r(y_{S})=0$. In
this case, not only must there exist a $k$-attractor at some $%
y_{D}<y_{K}<y_{S}$, but we know that it has {\it negative pressure} (since
$g(y_{K})<0$), is {\it stable} (since $w_{k}<1$, see Eq.~(\ref{rderivative}))
and is the {\it unique} $k$-attractor with negative pressure (since $%
r^{\prime }$ is monotonically decreasing in this $y-$interval).

Note also that this negative-pressure $k$-attractor only exists if there is
no dust tracker solution, that is, $r(y_{D})>1$. If there is a dust tracker,
($r\left( y_{D}\right) <1$), then, since $r^{\prime }(y)<0$ for $y>y_{D}$,
there is no point $y=y_{K}>y_{D}$ where $r(y)=1$ and, hence, there is no
$k$-attractor at $y_{D}<y<y_{S}$.

It is possible to have other  $k$-attractors with positive pressure
at $y<y_{D}$  (the closest one to $y_{D}$ is denoted by {\bf k}(?) in
Fig.1), but they will prove to be irrelevant in our scenario.

\section{Cosmic Evolution and Attractor Solutions}

Once all possible attractors for  
$k$-essence  have been identified, it is
easy to understand the evolution of the $k$-field as a voyage from one
attractor solution to another 
as different phases of cosmic evolution proceed.
For both the  radiation- and   matter-dominated phases,
there are several possible configurations of relevant attractor solutions.
In this section, we systematically  classify
the attractor configurations for each phase and their consequences 
for cosmic evolution.

\subsection{Radiation-Domination}

We assume that $g(y)$ has been chosen so that there exists
an attractor solution ({\bf R})
at $y=y_{R}$ such that $r^2(y_{R})\equiv \varepsilon _{k}/\varepsilon _{tot}$
is in the range one to ten percent, roughly equipartition conditions. This
energy ratio leads most naturally to a matter-dominated epoch that lasts a
few billion years and cosmic acceleration beginning at about the present
epoch. Depending on the form of $r^2(y)$, which is determined by 
$g(y)$ in the
Lagrangian,  there will be additional attractors
during the radiation epoch. Whether $y$ is drawn to the correct attractor $%
y_{R}$ depends on initial conditions and the other attractors. Ideally, we
want $y=y_{R}$ to have the largest basin of attraction so that most initial
conditions join onto the desired cosmic track. The combination of
cosmologically relevant attractors during the radiation-dominated phase can
be one of three types:

A$_{r}$) {\bf R, S }and {\bf no} other attractors at $y_{S}>y>y_{R}$. This
occurs only if the function $r\left( y\right) $ decreases for
$y_{R}<y<y_{S}$.
Conversely, if 
 $r\left( y\right) $  increases somewhere in the range  $%
y>y_{R}$ then it inevitably leads to the appearance of an 
extra $k$ and/or 
$r$ attractor at $y>y_{R}.$ Let us prove it.

If the function $r\left( y\right) $ increases within some interval,
it means
that the  derivative $r^{\prime }\left( y\right) $ is positive  there. On
the other hand, as it follows from (\ref{rderivative}), 
$r^{\prime } (y)$  is positive only if $w_k >1$.  
Since $w_{k}(y_{R})=1/3$ , $%
w_{k}(y_{S})=-1$ and $w_{k}(y)>1$ somewhere in the interval
$y_{R}<y<y_{D}$,  there
must be another point $\bar{y}$ within this interval, 
where $w_{k}(\bar{y})=1/3.$ If $%
r\left(\bar{ y}\right) <1,$ this point is a radiation tracker different from
{\bf \ R} with a different value of $r^2(y)$.
If $r^2(\bar{y}) >1$, then $\bar{y}$
is not a
tracker at all; 
 but,  since $r\left( y_{S}\right) =-1$,  there must 
exist a point in the interval $%
y_{S}>y_{k}$ $>\bar{y}$ 
where $r\left( y_{k}\right) =1,$ which corresponds to
a $k$-attractor.  That is, either there is a an extra radiation tracker or
there is an extra $k$-attractor.

For models of type A$_r$ where $r^2(y)$ is monotonically decreasing,
a dust tracker solution with $r(y_{D})<r(y_{R})$ is inevitable
and 
$k-$essence will be attracted immediately to it after matter-radiation
equality, a situation we are trying to avoid in order to
explain the present-day cosmic acceleration. The model $\widetilde{p}(X)=-1+X$
falls in the above category; with a field redefinition, the action can be
recast into the model of a field with canonical kinetic energy rolling down
an exponential potential \cite{expon}, an example which is well-known to
track in both the radiation- and matter-dominated epochs.

B$_{r}$) {\bf R, S, K }plus possibly other attractors at $y<y_{D}.$ This
situation takes place when there is no dust tracker solution ($r\left(
y_{D}\right) >1$) the case considered in our first paper.\cite{ArMuSt}

C$_{r}$) {\bf R, S (no K }attractor{\bf ) } and at least one additional
attractors {\bf r(?)} or {\bf k(?)}. This case occurs whenever there is a
dust tracker solution ($r\left( y_{D}\right) <1)$) with the property that $%
r\left( y_{D}\right) >r\left( y_{R}\right) $ or, in other words, $%
(\varepsilon _{k}/\varepsilon _{tot})_{D}>(\varepsilon _{k}/\varepsilon
_{tot})_{R}$. Even though there exists a dust tracker solution, we will show
it is  nevertheless
possible to have a finite period of cosmic acceleration at the present
epoch before $k$-essence reaches the dust tracker solution
in the future. 
For this to occur,
the function $r\left( y\right) $ 
must increase somewhere in the interval $y_{R}<y<y_{D}$. 
This is precisely the case considered above (see discussion of case
A$_r$), where we argued that there must be an extra $r$ and/or $k$-attractor
in the interval $y_D < y< y_R$.  Furthermore,
the attractor closest to $y_D$ must have 
 $r(y_{r/k})>r(y_{D})>r(y_{R})$; otherwise, we could find
 another attractor in the interval $y_{r/k} < y < y_k$,
as can be shown by repeating the argument 
presented under A$_r$ for this interval.
If $r(y_{r/k})>r(y_{D})>r(y_{R})$,
this second tracker has a larger fraction of $k$-essence.

\begin{figure}[tbp]
\begin{center}
  \epsfxsize=3.3 in \epsfbox{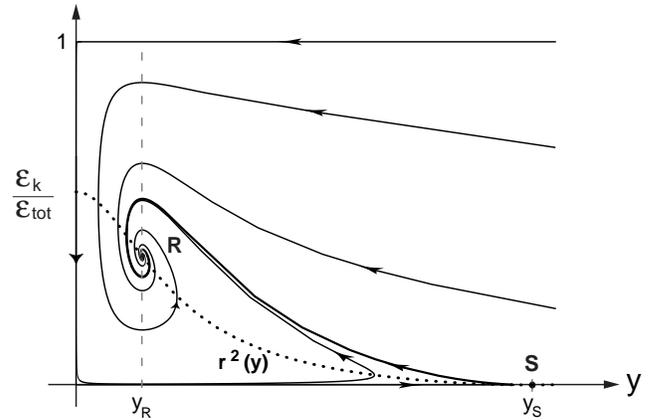}
 \end{center}
\caption{Phase diagram for case A$_r$ during the radiation-dominated epoch.
Phase lines flow in the direction shown by the arrows, dashed horizontal
lines determine the $y$ coordinate of attractor solutions and boldface
labels the corresponding attractor points. The dotted line shows the points
where $\protect\varepsilon_k/\protect\varepsilon_{tot}=r^2(y)$. }
\label{fig:Ar}
\end{figure}

\begin{figure}[tbp]
\begin{center}
 \epsfxsize=3.3 in  \epsfbox{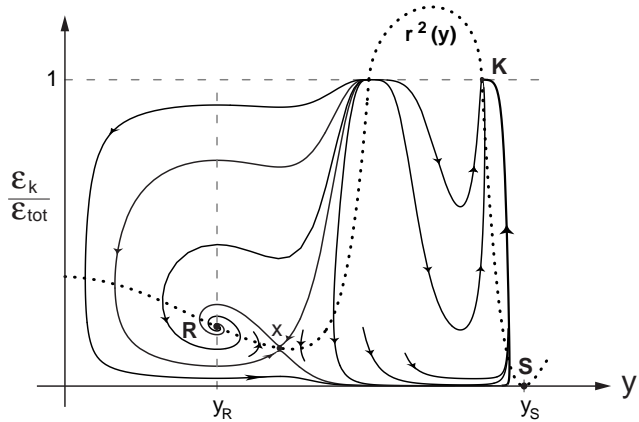}
 \end{center}
\caption{Phase diagram of a model of the type B$_r$ during the
radiation-dominated phase. In the relevant region of the diagram all
trajectories can be traced back to a common origin. Some of the phase
trajectories converge to the radiation tracker ${\bf R}$, while others,
after approaching the de Sitter point ${\bf S}$ finally reach the
${\bf K}$-attractor. The saddle point ${\bf x}$ ``separates'' both types of
trajectories.}
\label{fig:Br}
\end{figure}

\begin{figure}[tbp]
\begin{center}
  \epsfxsize=3.3 in   \epsfbox{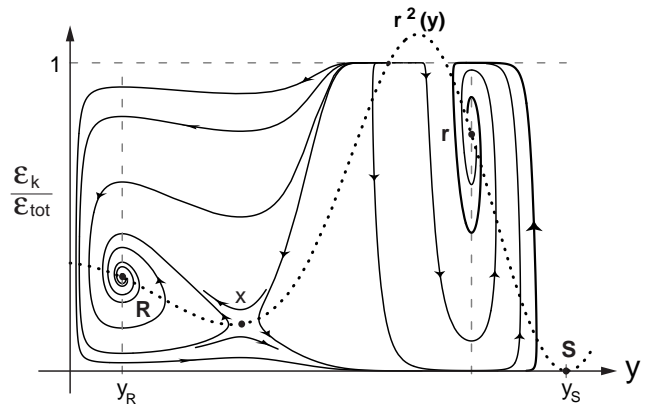}
 \end{center}
\caption{Phase diagram of a model of the type C$_r$ during radiation
domination, with same notation as in Fig. 3. }
\label{fig:Cr}
\end{figure}

A phase diagram of the system of Eqs.~(\ref{master1})-(\ref{master1a})
describing the global evolution of the $k$-field during radiation domination
is shown in Figs. \ref{fig:Ar}, \ref{fig:Br} and \ref{fig:Cr} for each of
the cases A$_{r}$, B$_{r}$ and C$_{r}$ respectively. Phase trajectories
cannot cross the lines where $\varepsilon _{k}/\varepsilon _{tot}$ is 
equal to zero or one, and, hence, their tangents are horizontal there.
The position of the radiation tracker {\bf R} is fixed by the intersection of
the $y=y_{R}$ line (dashed) and the $r^{2}(y)$ curve (dotted). If $r^{2}(y)$
is bigger than one at the intersection point, the tracker does not exist.
Notice that the phase trajectories go in the direction of increasing
(decreasing) $\varepsilon _{k}/\varepsilon _{tot}$ for $w_{k}(y)<1/3$
($w_{k}(y)>1/3$) and therefore, their tangents are horizontal at the points
where $w_{k}(y)=1/3$.
On the other hand, phase trajectories evolve in the direction of increasing
(decreasing) $y$ for $\varepsilon _{k}/\varepsilon _{tot}<r^{2}(y)$ ($%
\varepsilon _{k}/\varepsilon _{tot}>r^{2}(y)$) and at the points where these
phase lines cross the curve $r^{2}(y)$ their tangents are horizontal (see
Eq.~(\ref{eq:master})). The form of $r(y)$ also gives a clue about the
equation of state $w_{k}(y)$: in the region where $r\left( y\right) $ is
an 
increasing function of $y$ we have $w_{k}(y)>1$ and where it decreases
$w_{k}(y)<1$. Hence, as noted previously, $r(y)$ is what mainly determines
the structure of the phase diagram.

As clearly seen in the figures in all cases, if the $k$-field is initially
located near the {\bf R-}tracker, it converges to it. Therefore, the
basin of attraction is non-zero in all three cases. The attraction region
includes equipartition initial conditions, the most natural possibility.

For A$_{r}$, Fig. \ref{fig:Ar}, the {\bf R}-attractor has the largest basin
of attraction, the complete phase plane. If one starts, for instance, at $%
(\varepsilon _{k}/\varepsilon_{tot})_{i}=\exp \left( -30\right) (\varepsilon
_{k}/\varepsilon _{tot})_{R},$ then the $k$-field rapidly reaches the
vicinity of the de Sitter point {\bf S} and joins the attractor connecting
this point to the {\bf R-}tracker.

The cases B$_{r}$ and C$_{r}$ have limited basins of attraction, and so are
not as favorable from the point of view of initial conditions. If the energy
density of the $k$-field is much smaller  than the value at the
{\bf R-}tracker, the $k$-field travels first to the vicinity of the
{\bf S-}attractor, where it meets the phase trajectory that connects it to
the {\bf K}-attractor (case B$_{r}$) or the {\bf r}-attractor (case C$_{r}$).
In either situation, the field never reaches the {\bf R}-tracker. Although
the latter two  cases have smaller basins of attraction than case
A$_r $, only  cases B$_r$ and C$_r$
can produce cosmic acceleration today. One can simply assume that the
initial value of the $k$-field lies somewhere
in the basin of attraction, a reasonable
possibility. An alternative is to introduce additional $\varphi -$dependence
in the Lagrangian, as for instance, $L=g\left( y,\varphi \right) /y\varphi
^{2},$ where $g\left( y,\varphi \right) \rightarrow g_{1}\left( y\right)$
at high energies $(\varphi $ is smaller than some $\varphi _{0})$ and
$g\left( y,\varphi \right) \rightarrow g_{2}\left( y\right) $ at relatively
low energies $(\varphi $ is bigger than $\varphi _{0}),$ such that
$g_{1}\left( y\right) $ has an A$_{r}$-set of attractors and $g_{2}
\left(y\right) $ has a B$_{r}$/C$_{r}$-set of attractors. Note, that the
exact value of $\varphi _{0}$ is not important at all, we only have to be
sure that the transition from one regime to the other happens before
equipartition. Although  modifying the Lagrangian may seem more
complicated, it has the advantage that it removes nearly altogether
dependence on initial conditions.

\subsection{Matter Domination}

We have shown that it is possible to choose a wide range of models and
initial conditions for which the $k$-field converges to the {\bf R-}tracker
during the radiation-dominated epoch. 
The goal is  to 
produce a scenario in which $k$-essence overtakes the matter density
and induces cosmic acceleration today.  Yet,
the contribution of $k$-essence to the total energy density
must not spoil   big bang nucleosynthesis
or dominate over the matter density at the end of the radiation-dominated
epoch (see Sec. III.B). To satisfy these conditions, it typically
suffices  if the {\bf R}-tracker satisfies
\begin{equation}
(\varepsilon _{k}/\varepsilon _{tot})_{R}=r^{2}\left( y_{R}\right) =\alpha
\simeq 10^{-2} - 10^{-1}.  \label{errad}
\end{equation}

In this subsection, we study the evolution as the universe 
 enters the matter-dominated epoch and the $k$-field is forced
to leave the radiation tracker. In a dust dominated epoch the relevant
attractors can appear in the following two possible sets: A$_{d}$) {\bf S, K 
}and{\bf \ }B$_{d}$) {\bf S, D}.

\begin{figure}[tbp]
\begin{center}
 \epsfxsize=3.3 in  \epsfbox{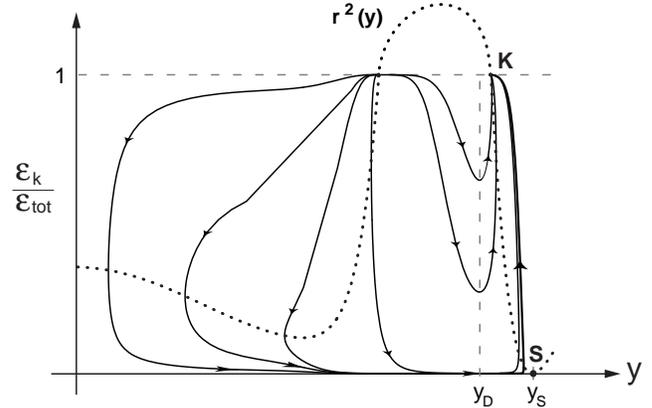}
 \end{center}
\caption{Phase diagram of a model of type A$_d$ during the matter-dominated
epoch. All trajectories have a common origin and all of them finally reach
the ${\bf K}$-tracker. Trajectories which ``skim'' the line $\protect%
\varepsilon_k/\protect\varepsilon_{tot}\approx 0$ reach this attractor after
going through a nearly de Sitter stage (the {\bf S}-attractor).}
\label{fig:Ad}
\end{figure}

In both cases successful $k$-essence models are possible. In the case A$_{d}$,
which was discussed in our earlier paper,\cite{ArDaMu} there is no dust
tracker solution, $(r\left( y_{D}\right) >1)$. Therefore, as seen in the
phase diagram of Fig. \ref{fig:Ad}, when the radiation-dominated epoch is
over, $k$-essence approaches first the {\bf S-}attractor;
afterwards,
when its energy density has increased significantly, it moves to 
the {\bf K-} attractor
(a state with negative pressure but $w_k>-1$). 
If $w_{k}\left( y_{K}\right) <-1/3$, the
expansion rate accelerates for the indefinite future; if $-1/3<w_{k}\left(
y_{K}\right) <0$, the expansion rate decelerates. Either way, the
matter-radiation density is increasingly negligible compared to $k$-essence
in the far future.

In the second case (B$_{d})$, there is a dust tracker solution. If $%
(\varepsilon _{k}/\varepsilon _{tot})_{D}\ll 1$,
 $k$-essence contributes
only a small fraction of the total energy density at this attractor, and
it approaches this attractor almost immediately after matter-radiation
equality.  This is not desirable since then $k$-essence cannot dominate
today or cause cosmic acceleration.  However,
if $(\varepsilon
 _{k}/\varepsilon _{tot})_{D}
=r^{2}\left( y_{D}\right) \rightarrow 1$ or $%
(\varepsilon _{k}/\varepsilon _{d})_{D}\gg 1$, there can be a 
period of cosmic acceleration before the $k$-field reaches the dust
attractor since it can first approach the {\bf S}-attractor and 
remain there for a finite time, see Figure~6.
Ultimately, though, the acceleration is temporary; the $k$-field
proceeds to the dust tracker, the expansion of the universe begins to 
decelerate, and the ordinary and (cold) dark matter density 
approaches a fixed, finite fraction of the total energy.
We refer to the scenario  as a ``late dust
tracker'' because the dust 
attractor is reached long after matter-domination has begun.

Taking into account that
$r(y_{D})$ is near unity or greater for both case A$_d$ and B$_d$,
we obtain from Eqs.~(\ref{definitionr}) and (\ref{errad}): 
\begin{equation}
\frac{g_{R}^{\prime }\,y_{R}^{2}}{g_{D}^{\prime }\,y_{D}^{2}}\leq \frac{9}
{16}\,\alpha \simeq 5\cdot (10^{-3}- 10^{-2}).  \label{r1}
\end{equation}
We can also infer from Fig.~1 that $g_{D}^{\prime }\cdot (y_{R}-y_{D})\leq
g\left( y_{R}\right) =-y_{R}\,g_{R}^{\prime }/3$ and, therefore, 
for  $\alpha \ll 1$, 
\begin{equation}
\frac{y_{R}}{y_{D}}\leq \frac{3}{16}\alpha \simeq 2\cdot (10^{-3}-
10^{-2}) 
\end{equation}
and
\begin{equation}
\frac{g_{D}^{\prime }}{g_{R}^{\prime }}\leq 
\frac{\alpha }{16}\simeq 6\cdot (10^{-4}- 10^{-3}).  \label{r2}
\end{equation}
Since $\varepsilon _{k}=-g^{\prime }/\varphi ^{2}$ and $\left| g^{\prime
}\left( y_{S}\right) \right| \leq \left| g^{\prime }\left( y_{D}\right)
\right| $, we conclude that after radiation domination, when the $k$-field
reaches the vicinity of the {\bf S}-attractor, the ratio of energy densities
in $k$-essence and dust can not exceed $\varepsilon _{k}/\varepsilon
_{d}<\alpha ^{2}/16\simeq 6\cdot (10^{-6}- 10^{-4}).$ This is the
nadir of $k$-essence; once $k$-essence approaches the {\bf S}-attractor, its
contribution to the cosmic density increases again until it becomes
comparable to the matter density. In case A$_{d}$, the $k$-field will evolve
further to the {\bf K-}attractor and the $k$-essence energy will
increasingly dominate over the matter density. In case B$_{d}$, the
$k$-field approaches the {\bf D}-tracker where the ratio of $k$-essence
to the matter  density
approaches some fixed positive value.

The statements above are generic and do not dependent significantly on the
concrete model as long as it satisfies the simple criteria formulated above.
Let us stress that the only {\bf ``}small{\bf ''} parameter used is the
ratio $(\varepsilon _{k}/\varepsilon _{tot})_{R},$ which has to be of the
order of $10^{-2}- 10^{-1},$ a very natural range for these models
and one that 
satisfies constraints of big bang nucleosynthesis (see Sec. III.B).
For this range, 
the present moment is approximately the earliest possible time when 
cosmic acceleration could occur.

\begin{figure}[tbp]
\begin{center}
 \epsfxsize=3.3 in  \epsfbox{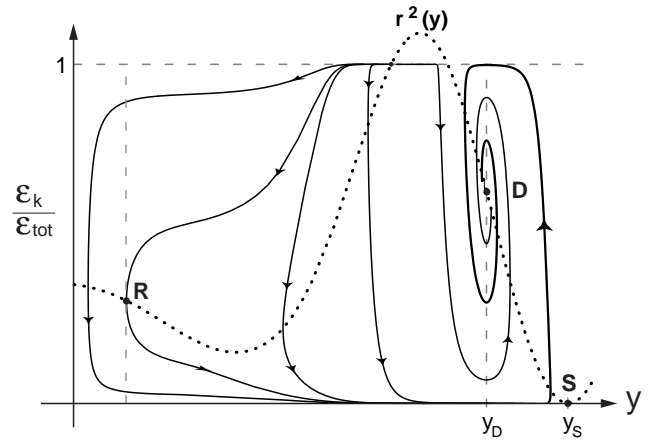}
 \end{center}
\caption{Phase diagram of a model of type B$_d$ during the matter-dominated
epoch. All trajectories have a common origin and all of them finally reach
the ${\bf D}$-tracker. Trajectories which ``skim'' the line $\protect%
\varepsilon_k/\protect\varepsilon_{tot}\approx 0$ reach this attractor after
going through a nearly de Sitter stage.}
\label{fig:Bd}
\end{figure}

Finally note that, during the transition from the radiation tracker 
{\bf R} to the de Sitter
attractor {\bf S}, 
the equation-of-state of $k$-essence has to take values bigger than
one, and hence the dominant energy condition $\varepsilon _{k}>|p_{k}|$ is
violated during a certain finite time interval. 
This violation
 implies that $k$-essence energy can travel with
superluminal speeds \cite{Wald}. Thus, perfectly Lorentz-invariant theories
containing non-standard kinetic terms seem to allow the presence of
superluminal speeds,
 as already pointed out in \cite{GaMu,GaMuOlVi}.

\section{Constructing Models}

In previous sections, we have presented a  general theoretical
treatment of the attractor behavior  of $k$-essence fields in a
cosmological background. We have emphasized the properties needed to
formulate models which will lead naturally to cosmic 
acceleration at the present epoch. In this section, we discuss how to
apply the general principles to construct
illustrative toy models.

Let us  summarize  the  conditions we have derived for building viable
Lagrangians. First, we must satisfy the general positive energy and
stability conditions in Eq.~(\ref{newenergy}). If $g$ takes positive and
negative values, they already suffice to guarantee generically the
existence of a radiation point $y_R$ where $w(y_R)=1/3$, a unique dust point
$y_D$ where $w(y_D)=0$, and a unique de Sitter point $y_S$ where $w(y_S)=-1$.
The radiation point is an attractor if $g''(y_R)$ is sufficiently small,
\begin{equation} \label{eq:R-stability}
  g''(y_R)<-4\frac{g'(y_R)}{y_R},
\end{equation}
and the remaining prerequisites needed to ensure a successful scenario are
then reduced to simple restrictions on the derivative of $g$  at two separate
values of y: 
\begin{enumerate}
\item[i)]At $y_R$, $r_R^2=-2 g'(y_R)y_R^2 \simeq 10^{-2}- 10^{-1}$.
\item[ii)]At $y_D$ either $r_D^2=-9\,y_D^2\,g'(y_D)/8>1$ or
$1-r_D^2=1+9\,y_D^2\,g'(y_D)/8\ll 1$.
\end{enumerate}
The first condition in ii) corresponds to cases where there is no
dust attractor, and the second condition to cases where there is a
dust attractor with a small matter to $k$-essence energy density ratio.

\begin{figure}
\begin{center}
 \epsfxsize=3.3 in   \epsfbox{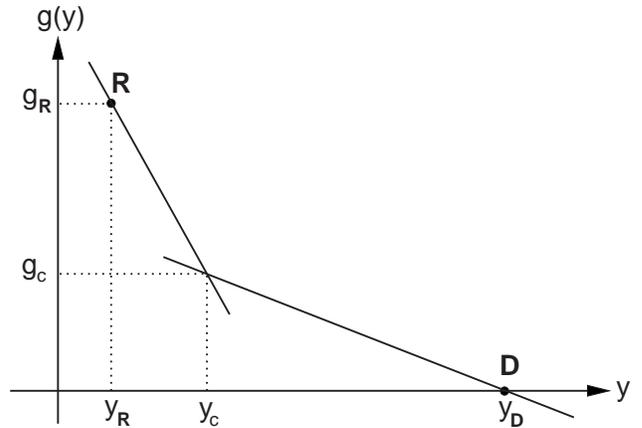}
 \end{center}
 \caption{
 A simple toy model for $g(y)$ consisting of two
 linear pieces  meeting at the ``crossing point'' $y_c$.
 Here $y_R$ and $y_D$ are the radiation and the dust attractor values,
 and the derivatives of $g$ at these points are
 $g'_R$ and $g'_D$, respectively.}
 \label{fig:glued}
 \end{figure}

A straightforward way of constructing a function with given derivatives at two
points is to glue two linear functions with the required slopes, as
shown in Figure \ref{fig:glued}. Observe that if $g(y)$ is linear around
the radiation point the attractor  requirement (\ref{eq:R-stability}) is
automatically fulfilled. In order to have a finite $c_S^2$,
it suffices to introduce small quadratic corrections to the
glued linear functions. We implement this procedure to build
a toy model expressed in terms of artificial parameters (from the 
point-of-view of fundamental physics) that can 
be simply related to Fig.~\ref{fig:glued} 
and our earlier discussion of attractor solutions.
One should appreciate that, for this pedagogical purpose, we 
have ``overparameterized" the problem --
the outcome is rather insensitive to most
parameters as long as they obey certain simple general
conditions.  Simpler forms with fewer parameters are 
certainly possible.

Let $g_{glue}(y)$ be any
smooth function constructed by gluing the two linear
pieces of Fig.~\ref{fig:glued}. The function $g_{glue}$ depends on $y$ and has
$y_R, g'_R, y_D$ and $g'_D$ as parameters where
$y_R$ and $y_D$ are the radiation and the dust attractor values 
and the derivatives of $g$ at these points are $g'_R$ and $g'_D$ respectively.
Our toy model corresponds to
\begin{equation}\label{interpol-g}
g(y)\equiv g_{glue}(y)
\left(1-\frac{y}{s^2\cdot y_D}\right).
\end{equation}
The factor
$g_{glue}$ describes the function in Figure \ref{fig:glued} and  the
factor in parenthesis provides the quadratic corrections needed to have a
positive speed of sound. It so happens that the latter 
factor also shifts the de Sitter point from $y=\infty$, 
as it would be for purely linear functions, 
to finite $y$, although this is not crucial for our purpose. 
For $s \gg 1$ the
de Sitter point is located at $y_S\approx s\cdot y_D$ and 
$g\approx g_{glue}$.

Once a general form for $g$ is known, such as the example above,
one can study how the model parameters affect the resulting cosmology.
Our conclusion is that the predictions of the toy model are relatively
insensitive to the gluing function or to 
the particular values of $y_R, y_D, g'_R, g'_D$ and $y_S$ as
long as they satisfy certain simple relations. 
For instance, what sets  the values of $\Omega_k$
and $w_k$ today? 
Do these  depend on the precise form of the interpolating function?
We have solved numerically the equations-of-motion for a wide range of 
gluing functions $g_{glue}$ in  Eq.~(\ref{interpol-g}). For a typical parameter
choice, the final value of $\Omega_k$ does not depend on the particular 
gluing function as long as $g_{glue}$ conforms closely enough 
to Fig.~\ref{fig:glued}.

The value of $\Omega_k$ today does  depend on the evolution of 
$\varepsilon_k/\varepsilon_m$.
At early
times the field is locked at the radiation tracker, and its fractional
energy density ratio is given by $-2 g'_R y_R^2$. After radiation-matter
equality the field can not follow the radiation tracker anymore and
its energy density  drops by several orders of magnitude
until $\varepsilon_k/\varepsilon_m$ reaches a minimum value
at the time $w_k$ falls below zero.
We shall label this  minimum value with the subscript ``min".
The energy density at this minimum is roughly given by
\begin{equation}\label{eq:ratio-nadir}
\left(\frac{\varepsilon_k}{\varepsilon_m}\right)_{min}\approx \,r_R^2
\frac{g'_D}{g'_R}.
\end{equation}
The position of the minimum in time only depends on the
distance between the radiation and crossing point $y_c-y_R$. 
As $y_c-y_R$ increases from zero,  
the minimum is shifted from matter-radiation equality
to later times.
After reaching the minimum, the field moves on to the de Sitter attractor 
and $\varepsilon_k/\varepsilon_m$ grows as $(z+1)^{-3}$, where $z$ 
is the red shift.  In order to have $k$-essence dominate today,
it must be that $\varepsilon_k/\varepsilon_m$ during the radiation
epoch lies roughly between $10^{-1}$ and $10^{-2}$.  Then, 
$(\varepsilon_k/\varepsilon_m)_{min}$ lies in the range
$10^{-4}- 10^{-6}$ and, provided $y_c$ is chosen appropriately, 
this has $k$-essence dominating at about the present epoch.  
One can see these conditions impose constraints on certain combinations
of our parameters, although in a fairly natural range not very
far from unity.

As discussed in Section IV.B, there are two possible future fates for
the universe depending upon whether there is a ``late dust tracker" 
solution or not.
By requiring $r_D^2>1$ we avoid a dust 
tracker and, therefore, insure that the $k$-field approaches the
$k$-attractor 
when $k$-essence starts to dominate. The equation of state of
$k$-essence at the $k$-attractor depends on the parameter $s$. By increasing
$s$ the equation of state $w_k$ at the $k$-attractor approaches $-1$, and in
the limit $s\to \infty$, $w_k(y_K)\to -1$.  If $w_k < -1/3$, the 
expansion rate of the universe accelerates forever.
Using the maximal value of the $w$ at the present epoch as  allowed by
supernovae observations, say, $s$  can be simply adjusted to insure 
that  $w$ at the $k$-attractor is less than or comparable to this value.
In this case, the  equation-of-state of $k$-essence today will be less than
or equal to $w_k(y_K)$, which is set by $s$, as described above. 

If $r_D^2<1$, it is possible to have successful 
models if $r_D^2$ is sufficiently close to $1$. In such a model the 
equation-of-state of $k$-essence will finally reach $w_k=0$ in the far 
future; so, ultimately, cosmic acceleration ceases and the expansion begins
to decelerate again.
Nevertheless, it is still  possible to have a finite period
in which the equation-of-state is negative and which includes the
present epoch. It is worth noting that
models without a dust attractor are more generic and natural, since they do
not require a special tuning of $r\left(y_{D}\right)$  to a value close but
smaller than unity at the dust point.
Below we illustrate examples of both types.

\subsection{ Model without dust attractor} 

\begin{figure}[h]
\begin{center}
 \leavevmode
 \epsfxsize=3.3 in  \epsfbox{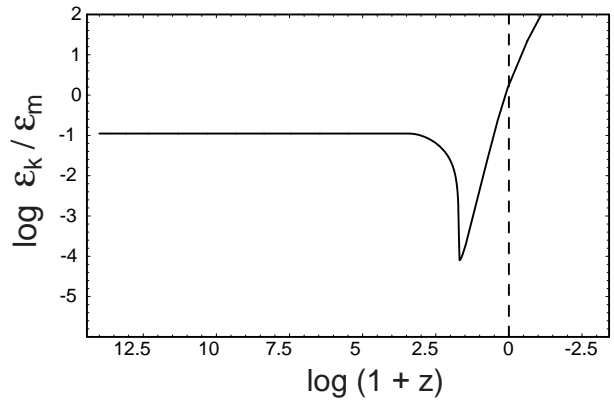}
\end{center}
\caption{\label{fig:nda-r} The ratio of $k$-essence to matter
energy density, $\varepsilon_k/\varepsilon_m$, 
vs. $1+z$ for a model with a $k$-attractor.}
\end{figure}

\begin{figure}[h]
\begin{center}
 \leavevmode
 \epsfxsize=3.3 in  \epsfbox{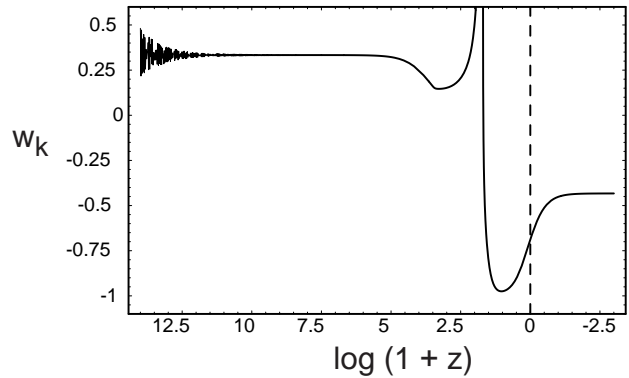}
\end{center}
\caption{\label{fig:nda-w}
The equation-of-state $w_k$ vs. $1+z$ for a model with a $k$-attractor.}
\end{figure}
Models that belong to the general class A$_d$  illustrated in Fig. 5 do
not have dust attractor solutions because $r(y_D)>1$.
Choosing the following values of the parameters,
$y_{R}=0.1$, $g_{R}^{\prime }=-5,$ $y_{D}=17$,
$g_{D}^{\prime }=-5\cdot 10^{-3}$ and $s^2 \cdot y_D=135$,  we have
$r(y_D)\approx 1.2$. Therefore, there has to be a {\bf K}-inflationary
attractor, which is located for our parameter choice  at $y_K\approx 28$.
At the {\bf K}-attractor, $k$-essence has the  equation-of-state
$w_{k}\left( y_{K}\right) \simeq -0.43.$ The ratio of the energy densities
at the {\bf R}-tracker in this model is
$(\varepsilon _{k}/\varepsilon _{tot})_{R}=0.1$. The results of the
numerical calculations are presented in Figs. \ref{fig:nda-r} and 
\ref{fig:nda-w}. We see that during the
radiation stage $k$-essence quickly reaches the radiation tracker, in
particular, the oscillations of the equation-of-state $w_{k}$ in Fig. 
\ref{fig:nda-w} around $w_k=1/3$ decay exponentially rapidly. The $k$-field
has the same equation of state as radiation until the moment when dust starts
to dominate. Around this time the energy density of $k$-essence suddenly drops
by three orders of magnitude and the equation-of-state, after a very short
period of increase, 
drops down to  $w_{k}\simeq -1$, the value of the equation-of-state along the
{\bf S}-attractor. After that, when  the energy density of $k$-essence becomes
significant, $w_{k}$ starts to increase towards the 
{\bf K}-attractor value, -0.43.  Since $\Omega_k$ is not yet unity,
the current value is somewhere between the {\bf K}-attractor value 
and -1; in this example, the value today ($z=0$) is $w_{k}\simeq -0.69$. 
The energy density of $k$-essence today is $\Omega_k\approx 0.65$,
and because we assumed a flat universe, $\Omega_m=0.35$. For completeness let
us mention that we have defined ``today'' ($z=0$) to be  the moment when the
matter-radiation energy density ratio is given by
$(\varepsilon_r/\varepsilon_m)_{today}
\equiv 4.307\cdot 10^{-5}/(\Omega_m h^2)$.

\subsection{Model with a late dust attractor}

\begin{figure}[h]
\begin{center}
 \leavevmode
 \epsfxsize=3.3 in  \epsfbox{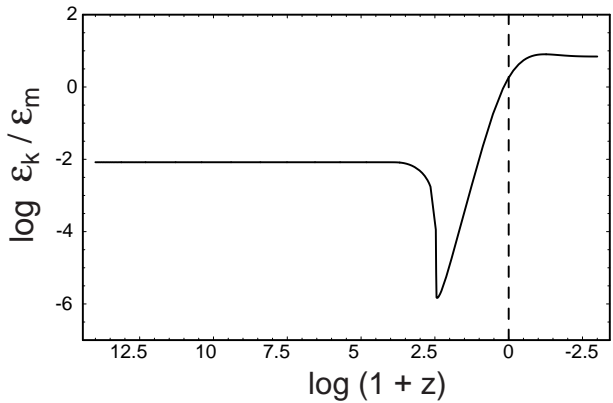}
\end{center}
\caption{\label{fig:lda-r}
The ratio of $k$-essence to matter
energy density, $\varepsilon_k/\varepsilon_m$,
vs. $1+z$ for a model with a late dust tracker solution. 
In this type of model, $w_k \rightarrow 0$ in the far future and
the ratio of $k$-essence to matter energy density approaches a constant.}
\end{figure}

\begin{figure}[h]
\begin{center}
 \leavevmode
 \epsfxsize=3.3 in  \epsfbox{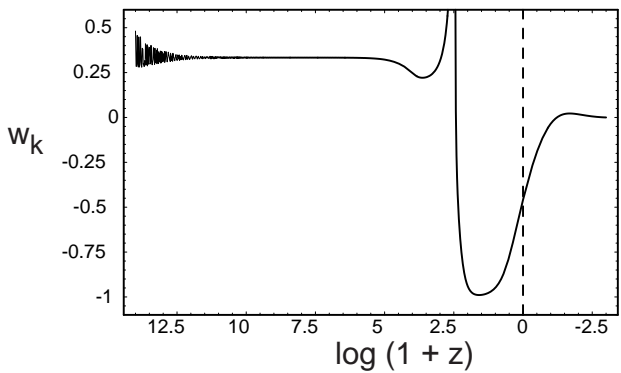}
\end{center}
\caption{\label{fig:lda-w}
The equation-of-state $w_k$ vs. $1+z$ for a model with a late dust tracker
solution.
}
\end{figure}

Taking  $y_{R}=11\cdot 10^{-3},$
$g_{R}^{\prime }=-34,$ $y_{D}=11,$ $g_{D}^{\prime }=-8\cdot10^{-3}$ and  
$s^2 \cdot y_D=56,$ we can construct a model with a 
``late dust  tracker", corresponding to the phase diagram in Fig.~6.
The parameters have been deliberately chosen to differ significantly
from the ones in the model without dust attractor in order to illustrate
that fine tuning is not necessary.
 
The late dust attractor is reached after $k$-essence passes near the 
de Sitter attractor following matter-radiation equality.
At the late dust tracker
$(\varepsilon _{k}/\varepsilon _{tot})_{D}=r^{2}\left( y_{D}\right)
\simeq 0.88$ and, correspondingly, $(\varepsilon _{k}/\varepsilon
_{d})_{D}\simeq 7.$ Hence, the fractional contribution of the
matter density is small but remains finite in the indefinite future.
The ratio of energies at the {\bf R-}tracker is $(\varepsilon
_{k}/\varepsilon _{tot})_{R}\simeq 8.3\cdot 10^{-3}.$  The results of  the
numerical calculations are presented in Figs. \ref{fig:lda-r} and
\ref{fig:lda-w}.
The evolution of the $k$-field here is very similar to the one we described
in the previous case; the differences  between both models occur at small
red-shifts. The fraction of the critical energy density of $k$-essence today
is in this model also $\Omega_k=0.65$  and the equation-of-state $w_{k}$
takes the value $-0.4$. The future
evolution of the model with a late dust attractor is completely different from
what we found in the previous one. Here the ratio of the energy densities
of $k$-essence and dust will continue growing in the future only until it 
becomes approximately $7$. After that it will start to oscillate around
this value with exponentially decaying amplitude  while the pressure
approaches the dust point, where $w_k=0$.

\subsection{Simpler and More Practical Examples}

The toy models presented thus far are all built
on the ansatz shown in Fig.~\ref{fig:glued}, which entails
numerous parameters.
We have pointed out that the large number of parameters 
is not a necessary feature.  We have introduced this form for
pedagogical purposes, since it enables one to 
study directly the relation between the attractor solutions and cosmic
evolution. Indeed,  our analysis showed that the cosmological solution is 
relatively insensitive to  most of the parameters provided they 
obey a few broad conditions.

To emphasize the point, consider a model of the form
\begin{equation}\label{simple-model}
\widetilde{p}(X) = -b + 2 \sqrt{1 + X\cdot h(a X)},
\end{equation}
where $h(a X)$ is  some smooth function that can be expanded in a 
power series in $X$.
This particular form is reminiscent of a Born-Infeld action
in which $h(a X)$ could 
represent  higher order corrections in $X$.
(This choice of a square-root form is not essential -- simply an 
example.)
As a specific case, for $b= -2.05$ and $X\cdot h(aX) = X - (a X)^2 + (a X)^3
- (a X)^4 + (a X)^5 - (a X/2)^6$ the Lagrangian defined by
(\ref{simple-model}) satisfies all constraints and produces
$\Omega_m= 0.3$ and $w_k=-0.8$ today if one chooses $a=10^{-4}$.  
This particular example has a cosmic evolution similar to the one 
described in Section V.A (no dust attractor).
We see that in this case, as  with
a wide range of other  functional forms,
the condition  $b>2 $   and the choice of 
the single parameter $a $ suffices to satisfy 
all of the conditions of the multi-parameter toy models. 

\section{Discussion}

Introducing a dark energy component with negative pressure has resolved many
observational problems with the standard cold dark matter model including
the recent evidence from supernovae searches that the universe is undergoing
cosmic acceleration. At the same time, the dark energy component presents a
profound challenge to cosmology and fundamental physics. What is its
composition and why has it become an important contribution to the energy
density of the universe only recently?

The example of $k$-essence shows that it is possible to find a predictive,
dynamical explanation that does not rely on coincidence or the anthropic
principle. Unlike a cosmological constant or quintessence models of the
past, the energy density today is not fixed by finely-tuning the vacuum
density or other model parameters. Rather, the energy density today is
forced to be comparable to the matter density today because of the dynamical
interaction between the $k$-essence field and the cosmological background.

Technically, the $k$-essence approach, at least in the examples we have
constructed, relies on attractor properties that naturally arise if the
action contains terms that depend non-linearly on the gradients of the
$k$-essence field. Non-linear terms of this type appear in most models unifying
gravity with other particle forces, including supergravity and superstring
models. In the past, these contributions have been ignored for reasons of
``simplicity.". The example of $k$-essence demonstrates that the effects of
non-linear dynamics can be dramatic. In a cosmological setting, we have
shown how they can cause the $k$-essence field to transform from a tracking
background field during a radiation-dominated epoch into a an effective
cosmological constant at the onset of matter-domination. This effect
explains naturally why cosmic acceleration could begin only at low
temperatures, at roughly the present epoch.

The non-linear dynamics is totally missed if the kinetic energy terms are
truncated at the lowest order contributions. Hence, the kinds of attractor
effects discussed in this paper have gone unnoticed in most treatments of
quantum field theory. This was one of the reasons for providing a detailed,
pedagogical treatment for at least one class of models. Clearly, this is the
tip of a broad arena of study. As another possible application, it is
interesting to note that a fundamental problem of superstring models is to
control the behavior of the many moduli fields in the theory, which are
coupled to one another through non-linear kinetic energy terms. At the
linear level, the moduli appear to be free fields with a flat potential, and
so there is no guidance as to why, amongst all the possible limits of
$M$-theory, the low energy limit looks like the Standard Model. Perhaps
non-linear attractor behavior constrains the evolution of moduli fields.

In this paper, we have focused on how non-linear dynamics addresses a
fundamental theoretical issue, the cosmic coincidence problem. An important
question to consider is whether there are observational tests to distinguish 
$k$-essence from alternative explanations. One notable feature of
$k$-essence models compared to the more general tracker quintessence
models\cite {ZlWaSt,StWaZl} is that the equation-of-state, $w_k$ is increasing
at the present epoch. For quintessence scalar fields rolling down tracker
potentials, the quintessence tracks the matter density ($w =0$) during most
of the matter-dominated epoch, and only recently has begun to decrease
towards $w=-1$. Hence, measurements of $dw/dz$ for the dark energy would
distinguish these two possibilities from one another and from a cosmological
constant. However, this test would not distinguish $k$-essence from more
general quintessence models that can also 
be tuned so that $w_k$ is increasing
today as well. A second feature of $k$-essence is the non-linear kinetic
energy contribution. A consequence is that the effective sound speed $c_S^2$
is generically different from unity, whereas $c_s =1$ for a scalar field
rolling down a potential. Depending on the model, the distinctive sound
speed can have subtle or significant effects on the cosmic microwave
background anisotropy. We will address these observational considerations in
a forthcoming paper.\cite{future}

As regards the future of the universe, our work here offers a new, 
perhaps  pleasing possibility. In previous models with cosmological
constant or quintessence, the acceleration of the universe continues 
forever and ordinary matter that composes stars, planets and life
as we know it becomes a rapidly shrinking fraction of the energy 
density of the universe.  In the ``late dust tracker" scenario
which we have introduced here, the acceleration is temporary and
the matter density approaches a fixed, finite fraction of the total.

This work was supported in part by the ``Sonderforschungsbereich 375-95
f\"ur Astro-Teilchenphysik" der Deutschen Forschungsgemeinschaft (C.A.P. \&
V.M.) and by Department of Energy grant DE-FG02-91ER40671 (Princeton)
(P.J.S.).

\end{document}